\documentclass[epj]{webofc}
\usepackage[utf8]{inputenc}
\usepackage[varg]{txfonts}   
\usepackage{booktabs}
\usepackage{xcolor}
\definecolor{darkred}{rgb}{0.4,0.0,0.0}
\definecolor{darkgreen}{rgb}{0.0,0.4,0.0}
\definecolor{darkblue}{rgb}{0.0,0.0,0.4}
\usepackage[bookmarks,linktocpage,colorlinks,
    linkcolor = darkred,
    urlcolor  = darkblue,
    citecolor = darkgreen]{hyperref}
\usepackage{subfigure}
\usepackage{slashed}
\usepackage{caption}
\wocname{EPJ Web of Conferences}
\woctitle{Lattice2017}
\newcommand{\forceindent}{\leavevmode{\parindent=1em\indent}}
\begin{document}
%
\selectlanguage{english}
\title{%
Calm Multi-Baryon Operators
}
\author{%
\firstname{Evan}            \lastname{Berkowitz}        \inst{1}\fnsep\thanks{Speaker, \email{e.berkowitz@fz-juelich.de}\\
Corresponding slides are available at \url{https://makondo.ugr.es/event/0/session/94/contribution/344}.}
                                                                        \and
\firstname{Amy}             \lastname{Nicholson}        \inst{2,3,4}    \and
\firstname{Chia Cheng}      \lastname{Chang}            \inst{3}        \and
\firstname{Enrico}          \lastname{Rinaldi}          \inst{5}        \and    \\
\firstname{M.A.}            \lastname{Clark}            \inst{6}        \and
\firstname{B\'{a}lint}      \lastname{Jo\'{o}}          \inst{7}        \and
\firstname{Thorsten}        \lastname{Kurth}            \inst{8}        \and
\firstname{Pavlos}          \lastname{Vranas}           \inst{9,3}      \and
\firstname{Andr\'{e}}       \lastname{Walker-Loud}      \inst{3,9}
}
\institute{%
Institut f\"{u}r Kernphysik and Institute for Advanced Simulation, Forschungszentrum J\"{u}lich     \and
Department of Physics, University of California, Berkeley                                           \and
Nuclear Science Division, Lawrence Berkeley National Laboratory                                     \and
Department of Physics and Astronomy, University of North Carolina, Chapel Hill                      \and
RIKEN-BNL Research Center, Brookhaven National Laboratory                                           \and
NVIDIA Corporation                                                                                  \and
Scientific Computing Group, Thomas Jefferson National Accelerator Facility                          \and
NERSC, Lawrence Berkeley National Laboratory                                                        \and
Nuclear and Chemical Sciences Division, Lawrence Livermore National Laboratory
}
\abstract{%
There are many outstanding problems in nuclear physics which require input and guidance from lattice QCD calculations of few baryons systems.
However, these calculations suffer from an exponentially bad signal-to-noise problem
which has prevented a controlled extrapolation to the physical point.
The variational method has been applied very successfully to two-meson systems, allowing for the extraction of the two-meson states very early in Euclidean time through the use of improved single hadron operators.
The sheer numerical cost of using the same techniques in two-baryon systems has so far been prohibitive.
We present an alternate strategy which offers some of the same advantages as the variational method while being significantly less numerically expensive.
We first use the Matrix Prony method to form an optimal linear combination of single baryon interpolating fields generated from the same source and different sink interpolating fields.
Very early in Euclidean time this optimal linear combination is numerically free of excited state contamination, so we coin it a \textit{calm baryon}.
This calm baryon operator is then used in the construction of the two-baryon correlation functions.

\forceindent
To test this method, we perform calculations on the WM/JLab iso-clover gauge configurations at the SU(3) flavor symmetric point with $m_\pi\sim800$~MeV --- the same configurations we have previously used for the calculation of two-nucleon correlation functions.
We observe the calm baryon significantly removes the excited state contamination from the two-nucleon correlation function to as early a time as the single-nucleon is improved, provided non-local (displaced nucleon) sources are used.
For the local two-nucleon correlation function (where both nucleons are created from the same space-time location) there is still improvement, but there is significant excited state contamination in the region the single calm baryon displays no excited state contamination.

}
\maketitle

\section{Introduction \label{sec:intro}}

Understanding how the wide variety of forms of matter arises from the underlying Standard Model is one of the central goals of nuclear physics.
Making quantitative connections between QCD and nuclei is key for probing experimentally inaccessible environments, such as the cores of neutron stars, as well as interpreting the current and planned experiments which utilize nuclear environments to search for rare processes arising from beyond the Standard Model physics.

Lattice QCD is currently our only tool for understanding nuclear physics from first principles with quantifiable systematic uncertainties.
However, the application of lattice QCD to multi-nucleon systems is notoriously difficult due to the infamous signal-to-noise (and related, sign) problem \cite{Parisi:1983ae,Lepage:1989hd}.
This noise problem increases exponentially with Euclidean time, limiting the time extents which may be reached in a typical lattice calculation.
To make matters worse, excited state splittings for multi-particle systems are typically much smaller than for single particles.
To reach the much longer time extents required to resolve these states using standard spectroscopic methods thus requires exponentially larger computational resources.

Recently, an exciting development in the study of the distributions of the magnitude and phase (and their evolution in Euclidean time) of lattice QCD correlation functions has made strides towards defeating the signal-to-noise problem in 
the long-time limit at the cost of introducting an additional extrapolation \cite{Wagman:2017xfh,Wagman:2016bam,wagman}. A more traditional approach to avoiding the signal-to-noise problem involves creating better interpolating operators, so that one may extract the desired state at earlier Euclidean time.
Considerable work has gone into the construction of two-nucleon operators, starting from reasonable single-nucleon interpolating fields, for various states with connections to infinite volume scattering states in assorted partial wave channels, as well as bound states~\cite{Beane:2009py,Yamazaki:2012hi,Beane:2012vq,Berkowitz:2015eaa}.

There has also been some discussion in the literature regarding the potential for so-called ``false plateaus" \cite{Iritani:2016jie,Iritani:2017rlk,aoki-lattice2017,Beane:2017edf}, in which the tuning of operators may cause near-cancellation between excited state contributions, leading to imperceptibly slow variation with time in the extracted energies, which is further masked by the noise associated with nucleons.
These issues are particularly troublesome for correlation functions which are not positive definite, due to the use of different operators at source and sink, which can lead to local minima in the effective mass that masquerade as the true ground state.
A commonly known example of this near cancellation can occur when looking at a ratio of two-body to one-body correlators to extract the energy splitting directly.
In this case, inelastic single-nucleon excited states contribute to both the two-body and one-body correlation functions, but with slightly different energies and wavefunction overlaps due to interactions.

These considerations make the study of two- (and later, multi-) nucleon operators of utmost importance in the advancement of the application of lattice QCD to nuclear physics.
One of the issues to be sorted out is where most effort towards improving these operators should be focused.
For example, overlap onto different low-energy elastic scattering excited states is largely achieved by varying the spatial positions of the two nucleons within the box.
On the other hand, reducing the overlap onto single-nucleon inelastic excited states requires tuning of the individual nucleon wavefunctions.

Surprisingly, studies comparing our one- and two-nucleon correlators have suggested that single-nucleon excited state contaminations dominate two-nucleon correlators at late Euclidean times.
While one would na\"ively expect the smaller energy splittings of the two-nucleon elastic excited states to be the most problematic, this observation seems to imply that projection onto the non-interacting two-body eigenstates (with some further tuning) sufficiently diagonalizes the interacting eigenstates, while simple variations of quark-model based single-nucleon operators do not adequately discriminate between the ground and first-excited states of the nucleon.%
\footnote{It should be noted that the potential method~\cite{HALQCD:2012aa,Aoki:2012tk} utilized by the HAL QCD collaboration relies upon the assumption that only elastic excited state contamination contributes at large time.
These findings seem to cast doubt on such assumption.}
This observation will be instrumental in the construction of improved two-nucleon operators to be discussed in the remainder of this work.

\section{Improving Multi-nucleon LQCD Interpolating Fields \label{sec:improved}}

The best technique for constructing a set of good multi-particle operators is the variational method based on the Generalized Eigenvalue Problem (GEVP) \cite{Blossier:2009kd}.
Here, a large basis of operators is constructed and used at both source and sink to create a matrix of correlators, which is then diagonalized to produce the best overlap onto the true eigenstates of the system.
This method has been used extensively and very successfully in the mesonic sector \cite{Dudek:2012xn,Dudek:2014qha,Wilson:2014cna,Wilson:2015dqa,Bulava:2016mks,Briceno:2016mjc,Moir:2016srx}.
One of the benefits of this method is the symmetric nature of the correlation functions, which renders the correlators positive definite and greatly reduces the potential for false plateaus.

In the baryonic sector, contractions are more costly and the number of propagators that must be calculated is in general much larger due to statistical noise, rendering variational methods for two-baryons prohibitively expensive thus far.
One of the biggest obstacles is the need for identical operators at source and sink, requiring momentum projection of the baryon interpolating operators, at least stochastically, at both ends.
Multi-nucleon correlators are generally only projected onto definite momentum at the sink, which is sufficient to eliminate overlap onto states with different total momentum.
These issues have led to investigations of alternative methods that capture some of the benefits of the variational methods without the need for a symmetric matrix of correlation functions.
One method which has been used successfully for nucleons (and multi-baryons) is known as the Matrix Prony method, and will be discussed in the next section.

\subsection{Matrix Prony \label{sec:mp}}
The Matrix Prony (MP) method was first used for analyzing lattice QCD correlation functions in Ref.~\cite{Beane:2009kya} and subsequently applied to two and three hadron systems in Refs.~\cite{Beane:2009gs,Torok:2009dg,Beane:2009py,Beane:2011iw}.
The great utility of the method stems from its simplicity to formulate and implement, as well as an ansatz which is imposed that is easy to verify or disprove.
Begin with a vector of correlation functions constructed from one source but multiple sinks, such as the smeared-smeared (SS) and point-smeared (PS) correlation functions which are commonly generated.
For simplicity of discussion, we will consider only two sinks, but the method in principle generalizes to many.
One assumes there is a transfer operator, acting upon the correlation functions
\begin{align}
y(t+\tau) &= \hat{T}(\tau) y(t)\, ,
&\textrm{with}&&
y(t) &= \begin{pmatrix} y_{SS}(t)\\ y_{PS}(t) \end{pmatrix}\, ,
\end{align}
where $\tau$ is a user chosen parameter which can be varied.
In order to solve for $\hat{T}(\tau)$, one takes an outer product with $y(t)$ and factorizes $\hat{T} = M^{-1} V$,
\begin{equation}
M y(t+\tau) y^T(t) = V y(t) y^T(t)\, .
\end{equation}
The following ansatz is then made: ``\textit{Over a time interval $t\in[t_i,t_f]$, only 2 states meaningfully contribute to $y_{SS}(t)$ and $y_{PS}(t)$.}''
The ansatz can be tested by summing this equation over the specified time interval.
Up to arbitrary normalizations, the solution is given by
\begin{align}
M &= \left[\sum_{t=t_i}^{t_f} y(t+\tau) y^T(t)\right]^{-1}\, ,&
V &= \left[\sum_{t=t_i}^{t_f} y(t) y^T(t)\right]^{-1}\, .&
\end{align}
The transfer operator can then be numerically constructed and used to solve the eigenvalue problem
\begin{align}
\hat{T}(\tau)\, q_\lambda &= (\lambda)^\tau q_\lambda\, ,
&\textrm{with}&&
\lambda &= e^{-E_\lambda}\, .
\end{align}
This provides a ``black-box'' method of constructing linear combinations of the original correlation functions designed to couple only to the ground state (the largest eigenvalue) or the first excited state (the smaller eigenvalue),
\begin{align}\label{eq:mp_gs}
q_0(t) &= c_{SS}^{(0)}\, y_{SS}(t) + c_{PS}^{(0)}\, y_{PS}(t)\, ,
&\textrm{and}&&
q_1(t) &= c_{SS}^{(1)}\, y_{SS}(t) + c_{PS}^{(1)}\, y_{PS}(t)\, .
\end{align}
If the ansatz is satisfied, over the time window specified, these two linear combinations of correlation functions will be largely free of excited state contamination, which can be readily checked by constructing the effective masses of these correlation functions and/or analyzing them with a single exponential fit.
To understand the systematic uncertainty associated with this method, the choice of $t_i$ and $t_f$ can be independently varied over a window where the input correlation functions appear to be dominated by just two states~\cite{Junnarkar:2013ac}.

With only different choices of smearing at the sink, we have observed that extending this method to three or more correlation functions makes the analysis unstable:
more noise was introduced through the linear combination of more than two correlation functions, offsetting any extra gain in early time;
beyond two solutions, the MP method can produce unphysical complex eigenvalues which arise from the numerical data as small oscillations in time that often appear with high statistics calculations.

\subsection{Calming the nucleon \label{sec:calm}}

The MP method has been used very successfully to analyze pairs of correlation functions generated with a variety of LQCD actions~\cite{Beane:2009kya,Beane:2009gs,Torok:2009dg,Beane:2009py,Beane:2011iw,Junnarkar:2013ac}.
However, instead of using MP after the multi-nucleon correlation functions are generated, we first perform a MP analysis of the single-nucleon correlation function to determine the optimal linear combination of S and P sinks that isolates the ground state nucleon as early in Euclidean time as possible.
This optimized linear combination can then be used in the construction of the baryon blocks which are used with the unified contraction algorithms to construct the two-nucleon correlation functions~\cite{Doi:2012xd,Detmold:2012eu},
$\mathcal{B}_{\text{calm}} = \sum_{n={\rm SS,PS}} c_n^{(0)} \mathcal{B}_n$.
We call these baryon blocks `calm', because they are less excited---in the sense that the excited-state contamination is dramatically reduced.
There are a few advantages to this method compared to previous strategies.
\begin{itemize}
\item The single-nucleon excited state contamination is significantly reduced, allowing for the analysis of the one and two-nucleon correlation functions at much smaller Euclidean times, where the stochastic noise is exponentially smaller.

\item The remaining excited state contamination observed in the two (or more) nucleon system after the initial MP time $t_i$ can be more reliably associated with the excited two-nucleon elastic scattering states of interest.
This allows for a subsequent simultaneous analysis of correlation functions corresponding to different two-nucleon spatial configurations designed to couple to the various elastic scattering states in a box, without running into the aforementioned issues that arise when trying to use more than two operators at once for MP.

\item It gives confidence in the analysis of the ratio of correlation functions designed to give the most precise determination of the energy splittings: with the dominant single-nucleon excited state removed, the near cancellation which occurs in numerator and denominator between the inelastic excited state contributions will disappear.

\item This method is substantially less expensive than the full variational method as it relies upon a smaller number of quark propagators, and many fewer contractions.
This method also offers a numerical savings over the more traditional method of computing two and more nucleon correlation functions: the standard approach requires the contractions to be computed for all the different choices of sink operators.
For values of the pion mass used in present day calculations, the contraction cost is a substantial fraction of the total cost of the calculation, often exceeding the cost of obtaining propagators.
Our new method requires the contractions to be computed only once with the optimized linear combination of sink operators.
\end{itemize}
We note that the NPLQCD collaboration has previously investigated the application of MP to two-nucleon correlation functions~\cite{Beane:2009py,Beane:2011iw} by constructing sinks with all possible different combinations of single-nucleon operator smearings~\cite{mjs_communication}, gaining benefits from the reduction of single-nucleon excited states, but possibly leading to the difficulties which can occur when trying to tune more than two operators. Only by imposing the selection of the single-nucleon MP combination that eliminates the first excited state explicitly \emph{before} constructing two-nucleon operators do we gain the full advantage of our method.
As we will show below, combining this method with spatially displaced two-nucleon operators~\cite{Berkowitz:2015eaa} to help reduce overlap onto the first elastic two-nucleon excited states reveals the full power of this method.

\subsection{Results \label{sec:results}}
\begin{figure}
\includegraphics[width=0.48\textwidth]{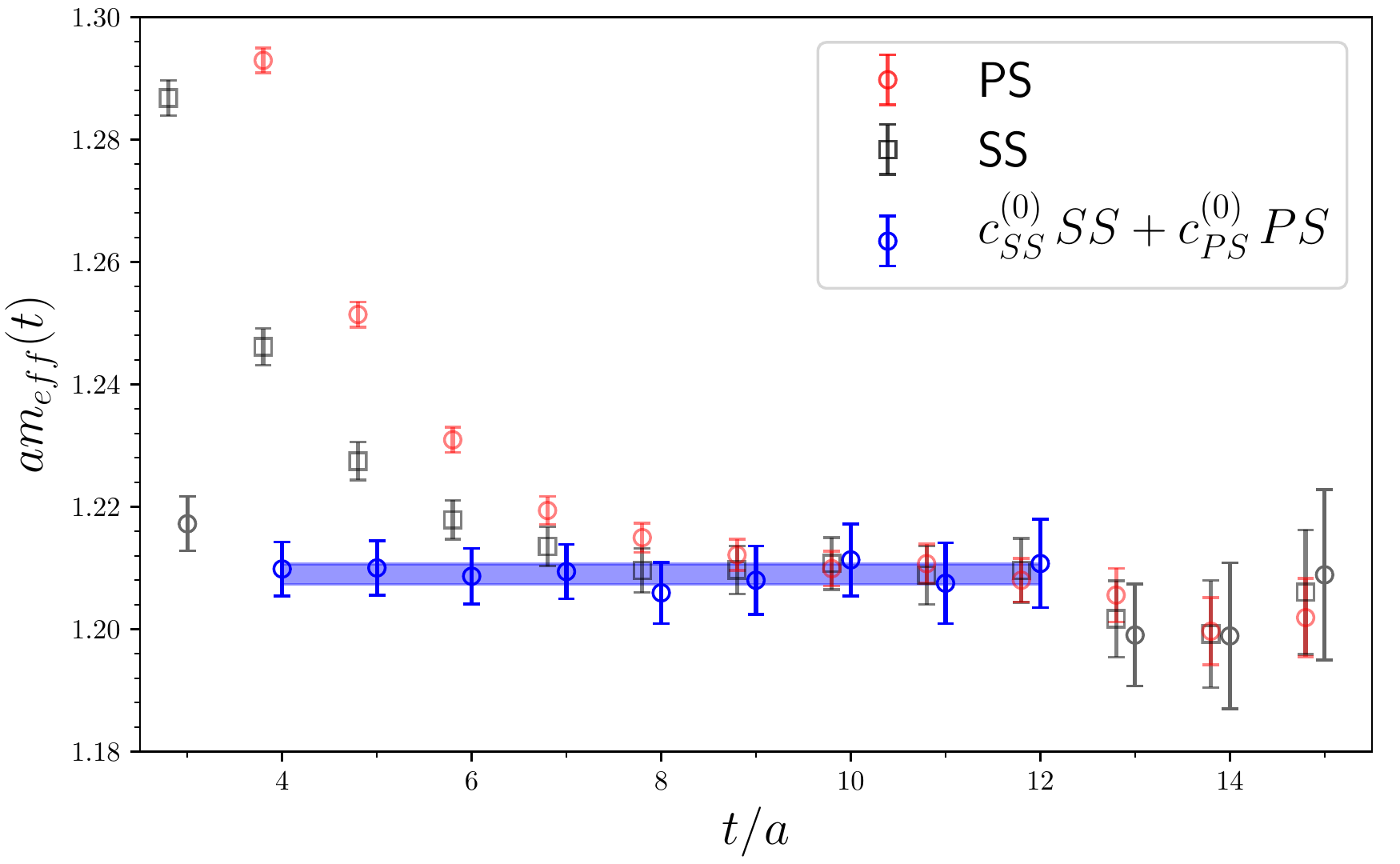}
\includegraphics[width=0.48\textwidth]{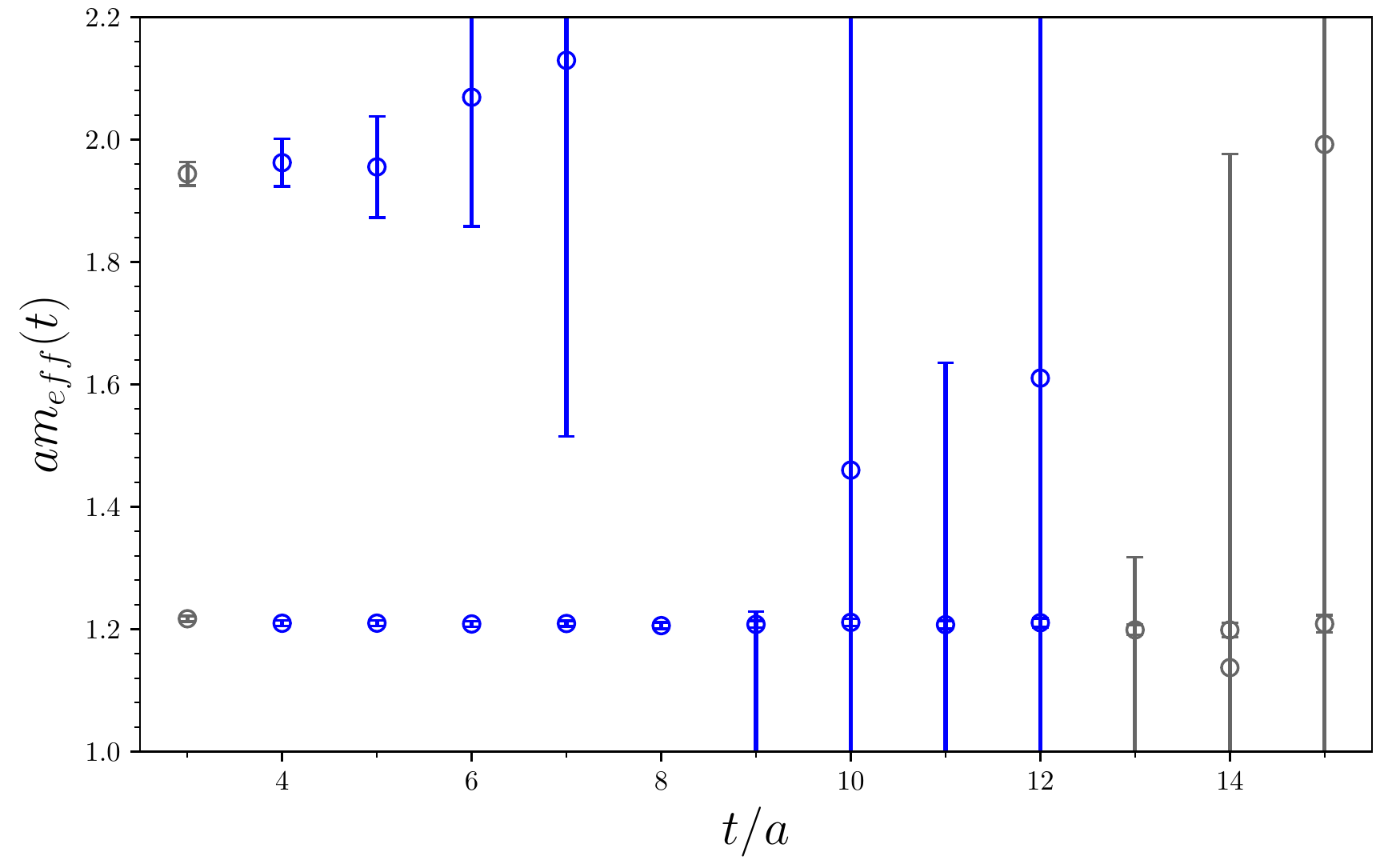}
\caption{\label{fig:nucleon}
(Left) The effective mass plot of the original SS and PS nucleon correlation functions as well as the optimized ground state correlation functions.
(Right) The effective mass of the two linear combinations of SS and PS correlation functions constructed from the MP analysis.
}
\end{figure}
As a first test of this new method, we apply it to the same set of gauge ensembles used in our previous calculation, Ref.~\cite{Berkowitz:2015eaa}, where we introduced the use of displaced nucleon operators at the source which were found to significantly improve the coupling to the ground states of interest
\footnote{
These configurations were generated by the WM/JLab group using an isotropic clover action, at the SU(3) flavor symmetric point with $m_\pi\sim800$~MeV and $a\sim0.145$~fm.
For more details, see Ref.~\cite{Beane:2012vq}.
} .
This allows us to provide a direct comparison with known results.
Specifically, for this comparison, we performed calculations on the smallest volume with $L/a = 24$, using a reduced set with one quarter the statistics.

In Fig.~\ref{fig:nucleon}, we show the effective mass of the nucleon generated from a point sink, a gaussian sink and optimized linear combination produced with MP, with a subset of 829 configurations.
As can be seen, the optimized ground state MP correlation function plateaus 5-6 time slices earlier than the original SS and PS correlation functions.

\begin{figure}
\includegraphics[width=0.48\textwidth]{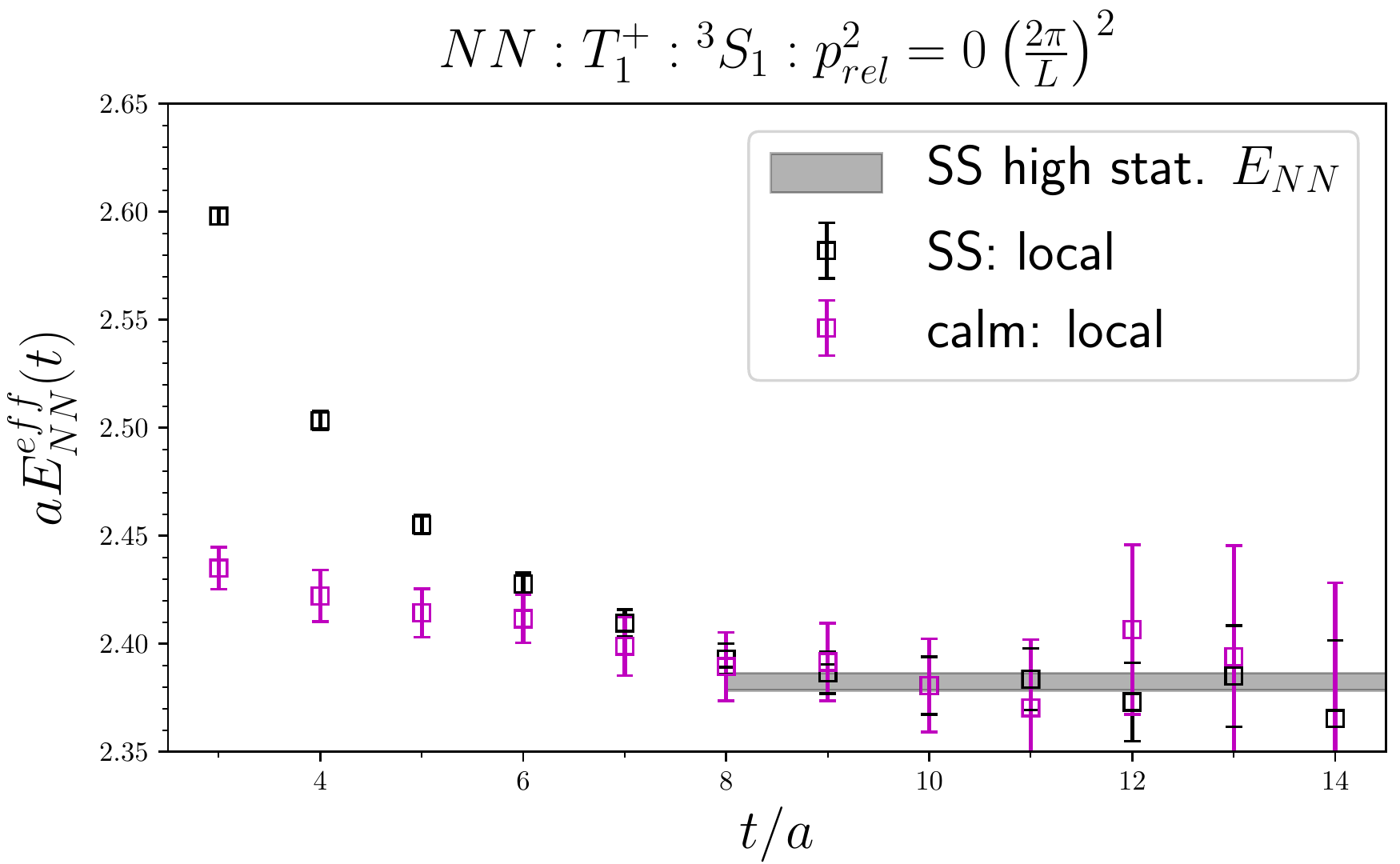}
\includegraphics[width=0.48\textwidth]{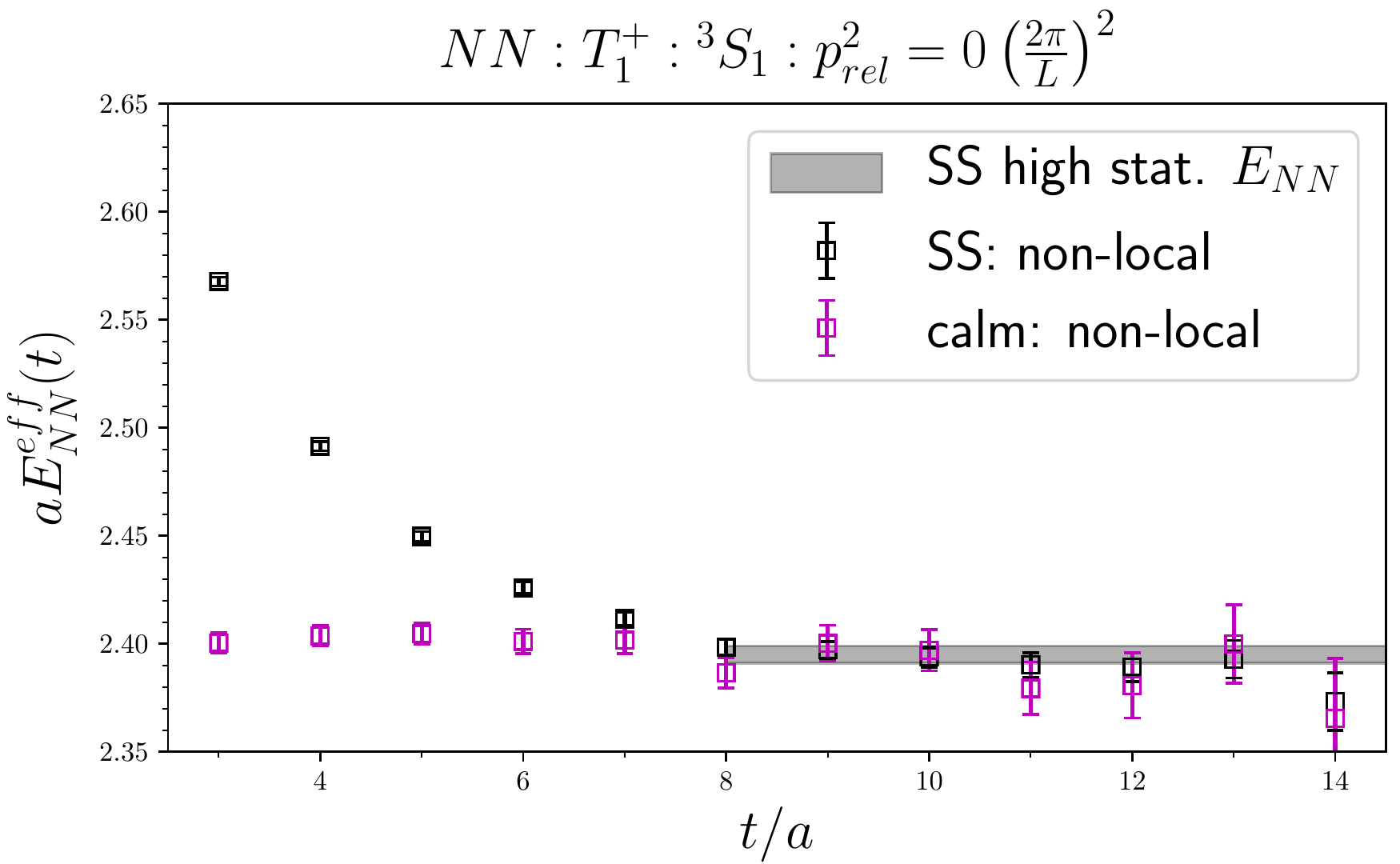}
\caption{\label{fig:nn_3s1}
The effective mass of the two-nucleon correlation functions constructed from the old SS single nucleon and the new calm-nucleon in the $T_1^+$ irrep which couples predominantly to the $^3S_1$ infinite volume channel.
(Left) Effective mass of the local interpolating operators at the source ($N^\dagger(0) N^\dagger(0)$).
(Right) Effective mass of the non-local interpolating operators at the source ($N^\dagger(0) N^\dagger(\Delta)$), introduced in Ref.~\cite{Berkowitz:2015eaa}.
}
\end{figure}

Let us now turn to the impact of using the calm nucleon in the two-nucleon correlation functions.
In Figure~\ref{fig:nn_3s1}, we show the effective mass of the two-nucleon correlation functions generated in the $T_1^+$ representation.
We show our original higher statistics results generated from the SS interpolating fields using both the local and non-local two-nucleon source operators.
In comparison, the two-nucleon operators are also constructed with the calm-nucleon operators.
In the left figure, we compare with the two-nucleon correlation functions constructed using the local interpolating field where both nucleons are created from the same spacetime point, $O^\dagger(t_0,0) = N^\dagger(t_0,0) N^\dagger(t_0,0)$.
As observed, the excited state contamination is noticeably reduced as compared with the SS interpolating field.
However, the stochastic noise is substantially larger than the original correlation function.

In contrast, in the right figure, we show the effective mass of the two-nucleon correlation functions with displaced operators at the source, first used in Ref.~\cite{Berkowitz:2015eaa}, $O^\dagger(t_0,0) = \sum_i c_i N^\dagger(t_0,0) N^\dagger(t_0,\Delta_i)$, where $\Delta_i$ are displacements of fixed length, summed over directions to overlap onto different cubic irreps.
We observe that the reduction of the excited state contamination in the two-nucleon correlation function is more significantly improved in this case, and more importantly, the stochastic noise is not much larger than the original correlation function.
This offers the promise of providing \textit{exponentially improved} two-nucleon correlation functions, as one will be able to analyze the correlation function starting much earlier in time, with minimal excited state contamination, in a region where the signal to noise is exponentially larger.

\begin{figure}
\includegraphics[width=0.48\textwidth]{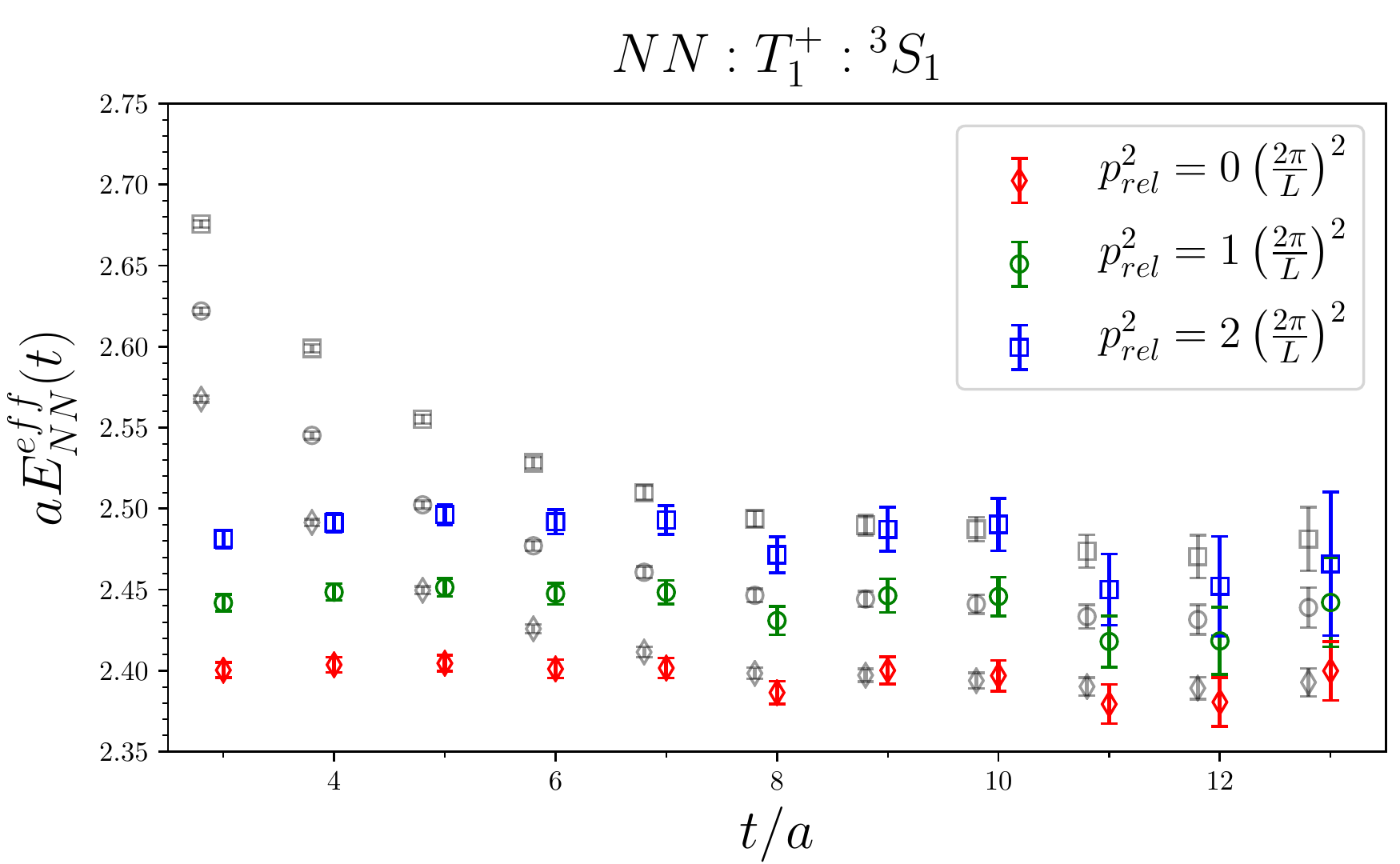}
\includegraphics[width=0.48\textwidth]{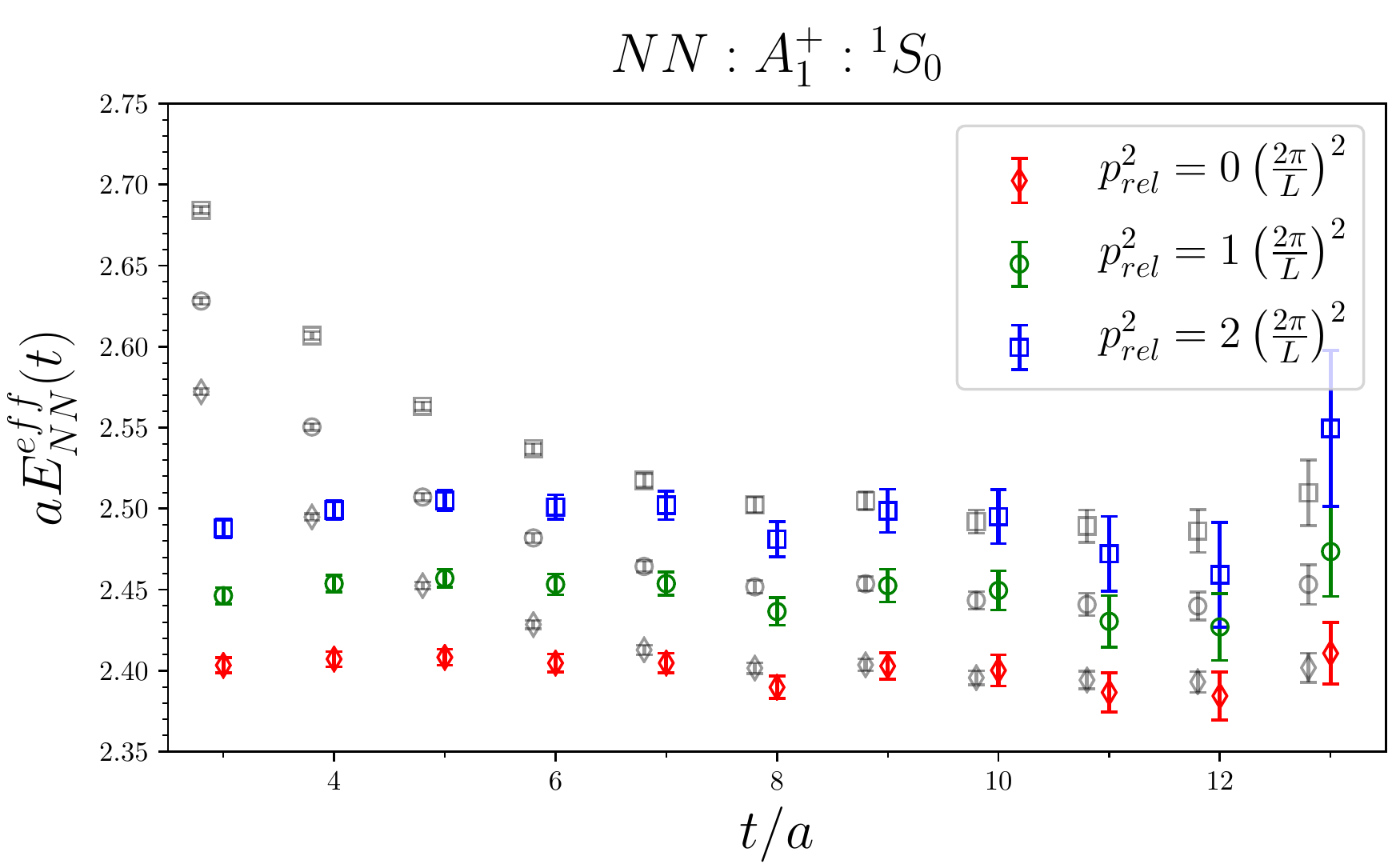}
\caption{\label{fig:nn_1s0}
Effective mass of the two nucleon system in the $T_1^+$ (left) and $A_1^+$ (right) irreps, with different colors representing calm-nucleon operators projected onto different relative momentum shells at the sink, and gray representing the old SS nucleons.
The calm nucleons were tuned with the $\mathbf{p}=0$ single nucleon operator.
}
\end{figure}

Finally, in Figure~\ref{fig:nn_1s0}, we show the same improvement is observed when the calm-nucleons are used in the first few non-zero momentum shells in which the individual nucleons are projected to back-to-back relative momentum $|\mathbf{p}| = |\mathbf{n}| 2\pi / L$, even when the calm-nucleons are determined from single nucleons with $\mathbf{p}=0$.
We find similar improvements in irreps that couple to the $p$-wave channels.

\section{Conclusion \label{sec:conclusion}}
The calm-baryon operators introduced in this work show a very promising ability to significantly reduce the excited state contamination in two-nucleon correlation functions.
The idea was motivated by
1) the need to extract meaningful information from two-nucleon correlation functions earlier in Euclidean time, before the stochastic noise overwhelms the signal,
2) the observation that the excited states in the two-nucleon system are dominated by the excited states of the single nucleon,
3) the realization that the Matrix Prony method can be used to construct a more optimal single nucleon sink operator which is numerically much less expensive that the full variational method.
The method was demonstrated on a set of configurations with a heavy pion mass and shows promise for being useful in all the cubic irreps we have studied in Ref.~\cite{Berkowitz:2015eaa}, including odd parity partial waves.
What will be particularly interesting is to test this method with a pion mass that is sufficiently light that reasonable extrapolations to the physical pion mass are feasible.
If the method proves as useful with light pion masses, the signal can be extracted approximately 1~fm earlier in Euclidean time where the signal-to-noise is exponentially larger, offering a numerically inexpensive \textit{exponential improvement}.
The method is readily applicable to multi-nucleon systems and channels with other flavors of baryons.

\section*{Acknowledgements}

We thank W. Detmold, R. Edwards, K. Orginos and D. Richards for use of the WM/JLab configurations used for this work.
The calculations were performed with \texttt{Chroma}~\cite{Edwards:2004sx} linked against \texttt{Quda}~\cite{Clark:2009wm,Babich:2011np} and managed with \texttt{METAQ}~\cite{Berkowitz:2017vcp}.
We thank R. Briceno and D. Wilson for useful conversations.
E.B. thanks M. Savage for further useful conversations regarding the prior development of similar techniques.
Numerical results were made possible through computing at Lawrence Livermore National Laboratory through the Multiprogrammatic and Institutional Computing and Grand Challenge programs, and through an ERCAP allocation at NERSC and an award of the Innovative and Novel Computational Impact on Theory and Experiment (INCITE) program on Titan to CalLat (2016).
This work was supported in part by the Department of Energy, Office of Science,
the RIKEN Special Postdoctoral Researcher Program and by the DFG and the NSFC Sino-German CRC110.

\bibliography{baryons}

\end{document}